\documentclass[preprint,showpacs,showkeys,amsmath,amssymb]{revtex4-1}
\usepackage{amsmath,amssymb,graphicx,multirow,setspace}
\date{\today}

\begin{document}


\title{Modified Li chains as atomic switches}

\author{Thomas Wunderlich$^1$, Berna Akgenc$^{1,2}$,
Ulrich Eckern$^{1,}$\footnote{ulrich.eckern@physik.uni-augsburg.de},
Cosima Schuster$^1$, and Udo Schwingenschl\"ogl$^3$}
\affiliation{$^1$Institut f\"ur Physik, Universit\"at Augsburg, 86135 Augsburg, Germany, \\
$^2$Kirklareli University, Physics Department, Kavakli, Kirklareli, Turkey\\
$^3$KAUST, PSE Division, Thuwal 23955-6900, Kingdom of Saudi Arabia}

\begin{abstract}
We present electronic structure and transport calculations for hydrogen and lithium 
chains, using density functional theory and scattering theory on the Green's function level,
to systematically study impurity effects on the transmission coefficient.
To this end we address various impurity configurations. Tight-binding results allow us
to interpret our the findings. We analyze under which
circumstances impurities lead to level splitting and/or can be used to switch between
metallic and insulating states. We also address the effects of strongly electronegative
impurities.
\end{abstract}

\maketitle

\section{Introduction}

Electronic devices have been reduced more and more in size over the last decades. Furthermore
it is now possible to place atoms or molecules accurately between macroscopic electrodes, 
hence experimental studies of the electronic transport for single atoms
\cite{scheer,Ha}, molecules \cite{molecules}, and nanowires \cite{wires} have become available.
One-dimensional structures are of particular interest due to their restricted transport channels,
making them prototypical model systems. For example,
mono-atomic chains have been realized by molecular
beam epitaxy \cite{scheer}. Using the tip of a scanning tunneling microscope, it has been
possible to place a row of eight atoms on a NiAl substrate \cite{nazin}.

From the theoretical
point of view, transport through distorted one-dimensional systems has been addressed already
in the early 1990ies by bosonization techniques \cite{Kane,Safi,Sassetti}. Transport
properties of lattice models are currently investigated on several levels. Comparison of an 
exact treatment by the density matrix renormalization group \cite{schmitteckert} with
approaches using density functional theory \cite{schenk} shows that the latter approach
often is sufficient to calculate the linear conductance, at least qualitatively.
The generalization of density functional theory to time-dependent potentials \cite{runge}
allows to study the propagation in time of the electronic states \cite{kurth}. 
To model the experimental setup more realistically, several methods 
have been developed for describing the
transmission through nano-contacts. Tight-binding formulations have been applied to
metallic nano-contacts \cite{chico,cuevas}, adding orbital information from chemical analysis;
however, these may fail in the contact regime. Nowadays, most apporaches rely on a
combination of density functional theory and a scattering approach on the non-equilibrium
Green's function level, based on the Landauer--B\"uttiker scheme \cite{Buttiker85}.

In this article, the electronic structure and the transmission coefficient are determined
for H and Li chains with defects. In Sec.\ II we present details
of the calculational method and the
structural setup, and discuss in Sec.\ III some results for H chains and their dependence
on the calculational parameters. In Sec.\ IV we turn to impurities in Li chains, focusing
on the interrelation between the energy level spectrum of the scattering region
and the transport properties.

\section{Computational details and structural setup}

\begin{figure}
\includegraphics[width=0.5\textwidth]{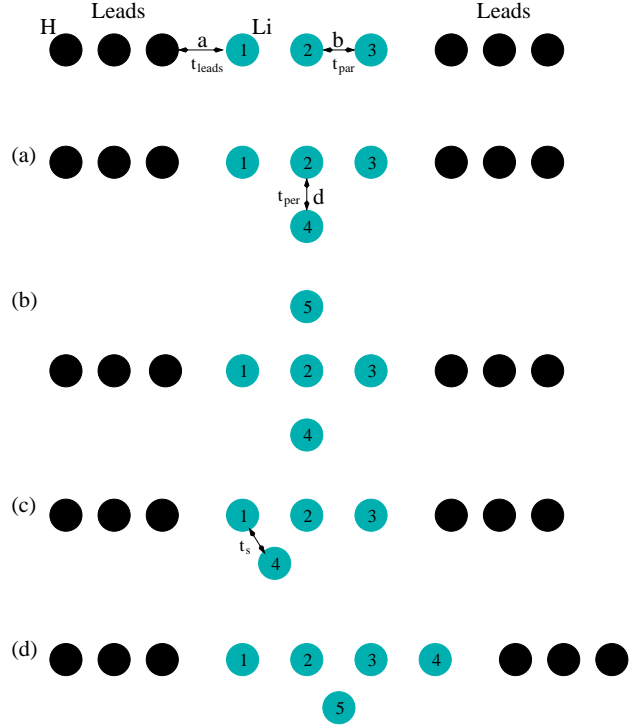}
\caption{Ad-atom configurations under consideration. The distance $d$ between the ad-atom(s)
and the chain is varied. The $t$-parameters refer to the tight-binding description of the
central scattering region; see main text.}
\label{figstruct_asym}
\end{figure}

We calculate the transmission through nanowires using a combination
of scattering theory and density functional theory \cite{Roch1,Roch2}, based on the
SMEAGOL and SIESTA \cite{Sol} codes. A single zeta basis set and the local density
approximation for the exchange-correlation potential are used. 
In order to model the experimental situation (without gate voltage), metallic leads
are connected to a central scattering region. In our case, the leads are H chains,
see Fig.\ \ref{figstruct_asym}. In the Landauer--B\"uttiker formalism, the
self-energies of the left (L) and right (R) leads are calculated first. The screening within the
metallic leads ensures that effects of the contact region decay within a few
nanometers. Since the leads are connected to the central scattering region (molecule,
nano-contact, or interface), an effective description of the central region (C)
emerges which includes the properties of the leads. In linear response,
the transmission coefficient is given by the retarded Green's function
$G_C$ of the central region and the lead self energies $\Sigma_{L/R}$. With
$\Gamma_{L/R}={\rm i}[\Sigma_{L/R}(E)-\Sigma_{L/R}^{\dagger}(E)]$ we have \cite{Meir92}
\begin{equation}
\label{eq6}
T(E,V=0)={\rm Tr}[\Gamma_L G_{C}^{\dagger}\Gamma_R G_{C}] \; .
\end{equation}
The conductance is given by $ G=\frac{2e^2}{h}T(E_F) $, where the
factor 2 accounts for the spin degeneracy \cite{Buttiker85}.

The impurity models studied in Sec.\ IV are displayed in Fig.\ \ref{figstruct_asym}.
The parent structure is a finite (metallic) Li chain with three atoms, which 
is coupled to H-chain leads. By varying the distance between the last H and the first Li
atom, we change the coupling from strong ($a=2.8$ \AA) to weak ($a=4.0$ \AA).
The distance between the Li atoms within the chain is fixed to $b=3$ \AA.
The impurities are Li atoms adjacent to the parent chain, thus breaking the rotational and/or
inversion symmetry of the system. In configuration (a) an additional Li atom is placed
directly below a chain atom, compare \cite{Cha07}. The distance $d$ between the ad-atom
and the chain is varied. A second atom on the other side of the chain, configuration (b),
restores the symmetry. An additional atom can also be placed
below a Li-Li bond, configuration (c), which also breaks the symmetry in the transport direction. 
In addition, we study configuration (d) in which we place the ad-atom below the central bond,
to obtain a system with inversion symmetry. 

As the transmission coefficient shows resonance peaks at the energy levels of the scattering
region, an analysis of this spectrum allows an explanation of the behavior of $T(E)$. 
For this purpose, the energy levels can be discussed on the basis of a tight-binding approximation 
with nearest-neighbor hopping. With the hopping parameters indicated in
Fig.\ \ref{figstruct_asym}, the Hamiltonian for a three-atom Li chain is given by
\begin{equation}
H = -t_{\rm par}\left(c^+_1c_2+c^+_2c_3+ {\rm h.c.}\right),
\end{equation}
where the sites are numbered 1, 2, and 3. The hopping energy along the chain, 
$t_{\rm par}$, is of the order of 2 eV, as the bandwidth of a Li chain with atomic
spacing $b=3$ \AA\ is $\sim 4.5$ eV. The three-atom chain has the energy levels
$\varepsilon_{1,3}=\pm \sqrt{2}t_{\rm par}$, and $\varepsilon_{2}=0$.
The coupling to the leads can be taken into account by an additional
hopping to the left and to the right of the scattering region.
The hopping to the ad-atoms is denoted by
$t_{\rm per}$, thus $H_{\rm per}=-t_{\rm per}(c_2^+ c_4 + {\rm h.c.})$ for configuration
(a), and $H_{\rm per}=-t_{\rm per}(c_2^+ c_4 + c_2^+ c_5 + {\rm h.c.})$ for (b).
In case of configurations (c) and (d), the hopping terms are given by
$-t_{\rm s}(c_1^+c_4+c_2^+c_4 + {\rm h.c.})$, and
$-t_{\rm s}(c_2^+c_5+c_3^+c_5 + {\rm h.c.})$, respectively.

\section{Hydrogen chains}

\begin{figure}[t] 
\begin{center}
\includegraphics[width=0.7\textwidth,clip]{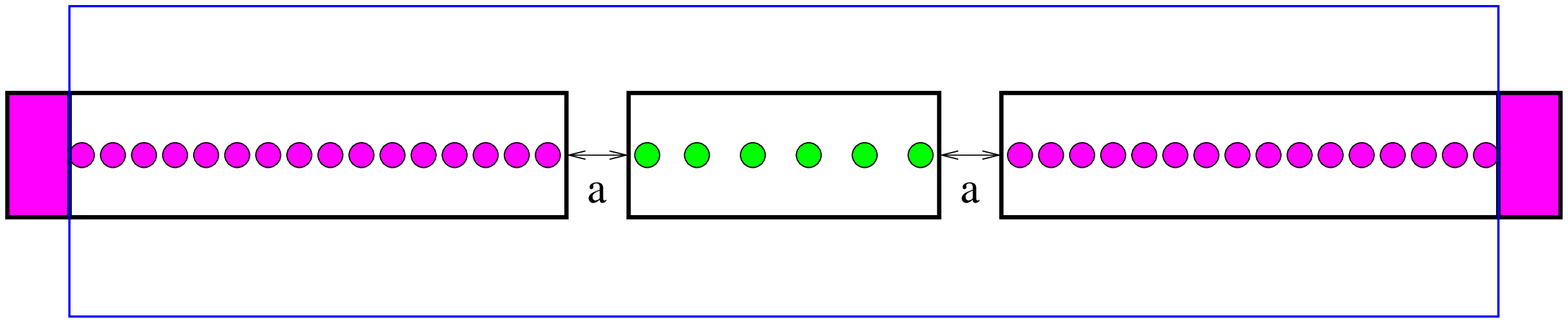}\\[0.5cm]
\includegraphics[width=0.45\textwidth,clip]{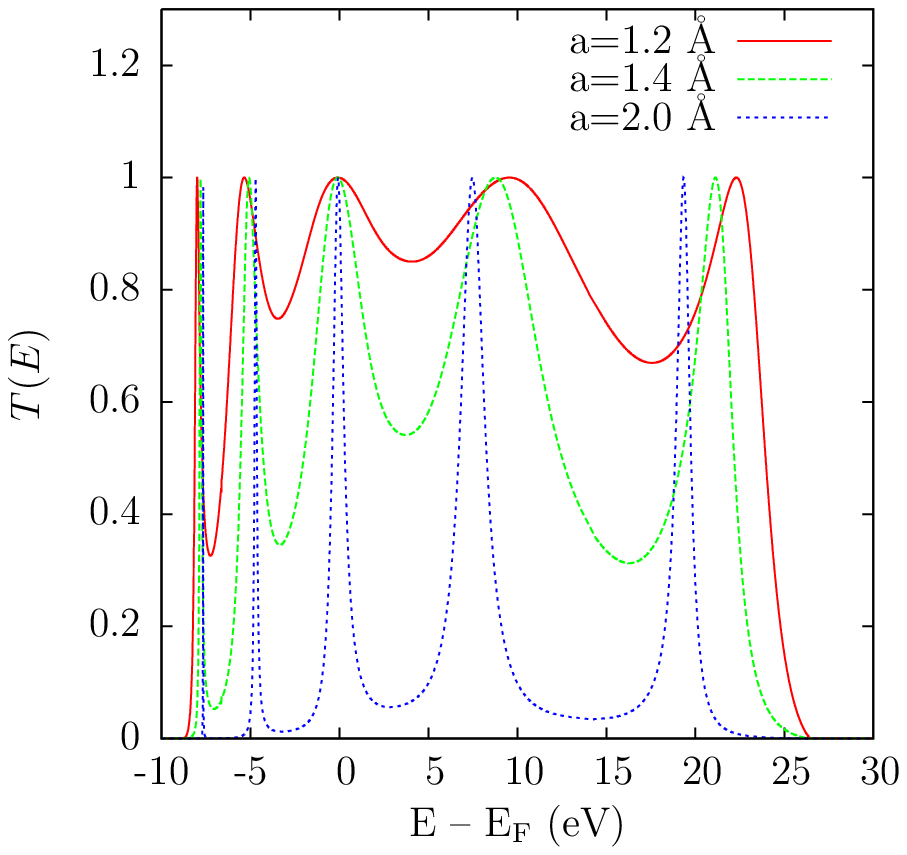}
\includegraphics[width=0.45\textwidth,clip]{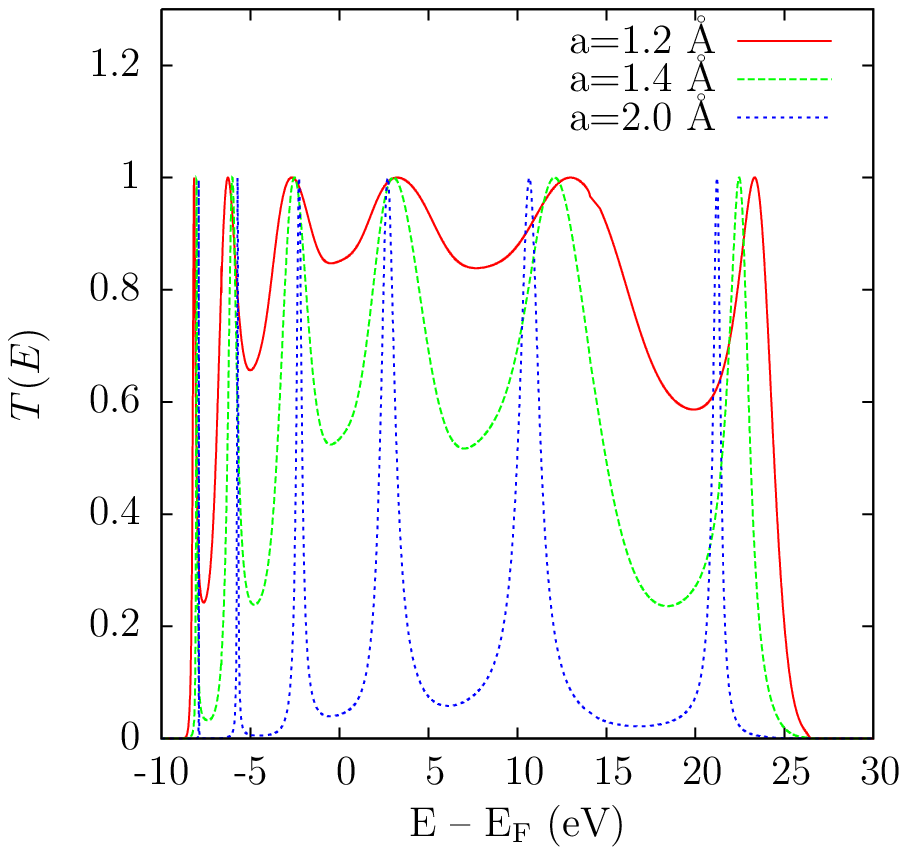}
\end{center}
\caption{Transmission coefficients of five-atom (left) and six-atom (right) H chains coupled
to H leads, upon increasing
the distance $a$ between the scattering region and the leads.
The bond lengths in the scattering region and in the leads are fixed at 1.0 \AA.}        
\label{fig2}  
\end{figure}

We start our discussion with a review of results for H chains, which is
the simplest case of one-dimensional scatterers, see the structural setup
in Fig.\ \ref{fig2}. He and coworkers \cite{He07} have used a system of infinite H
chains as leads and a finite six-atom H chain for their transport calculations, focusing
on the influence of the exchange-correlation potential and 
the according self-interaction errors. In the following we use the local density
approximation, since it reproduces well the qualitative behavior of the transmission
coefficient \cite{schenk}. Clearly, $T(E)$ depends on the energy levels of the
central H chain, and on the coupling strengths to the leads. The number of atoms within the chain
determines the number of energy levels, and hence the number of resonances, see Fig.\ \ref{fig2}.  
A five-atom chain is found to be metallic, whereas a six-atom chain turns out to be
insulating. This odd-even behavior is reflected by a maximum or minimum in $T(E_F)$
in Fig.\ \ref{fig2}. Moreover, the transmission peaks become 
broader with increasing coupling to the leads, and also shift in energy.

Considering in addition the bond length within the chain, $b$, as a parameter,
we find, as expected, a decrease of the band width with increasing $b$.
For $b=1.0$ \AA\ the bands are close to parabolic, and the unoccupied levels are more
sensitive when decreasing the coupling than the occupied levels.
With increasing $b$, the dispersion becomes more and more symmetric and cosine shaped,
consistent with the tight-binding description \cite{Diplomarbeit}. Keeping the lead-chain
distance fixed at, say, $a=1.2$ \AA\, but increasing the bond length $b$ beyond this
value, one arrives at a situation where the first and the last atom of the central region
are effectively bound to the leads, i.e., the scattering region is a diluted chain,
however, with two atoms less \cite{Diplomarbeit,note}.

\section{Lithium chains}

\begin{figure}
\includegraphics[width=0.45\textwidth,clip]{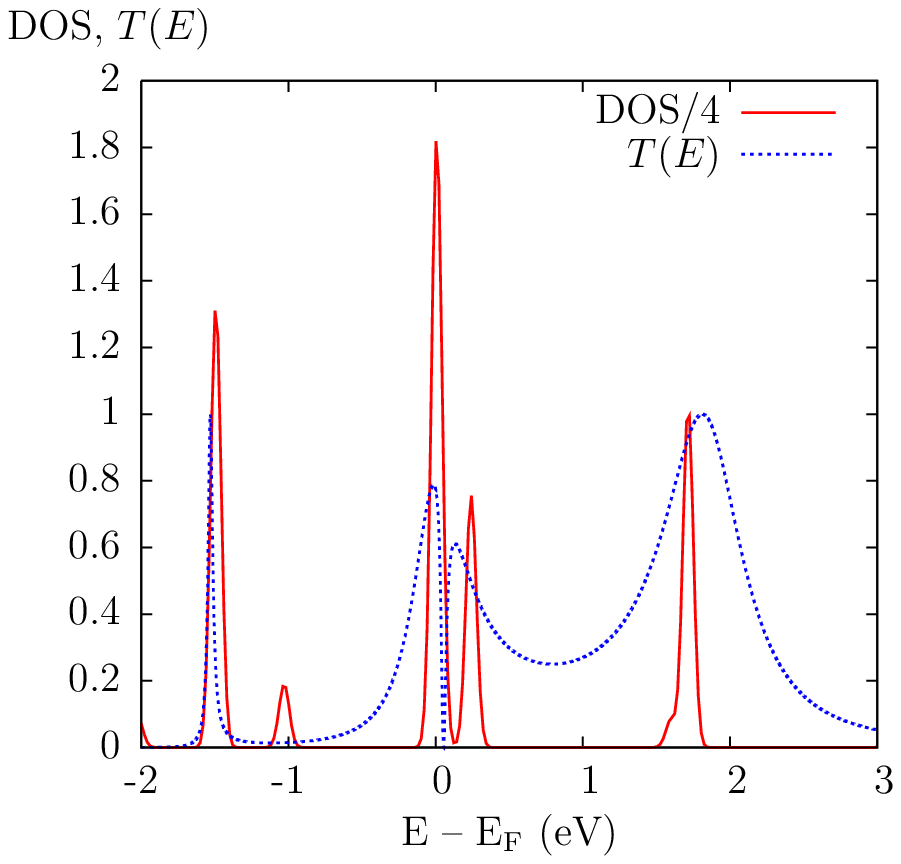}(a)
\includegraphics[width=0.45\textwidth,clip]{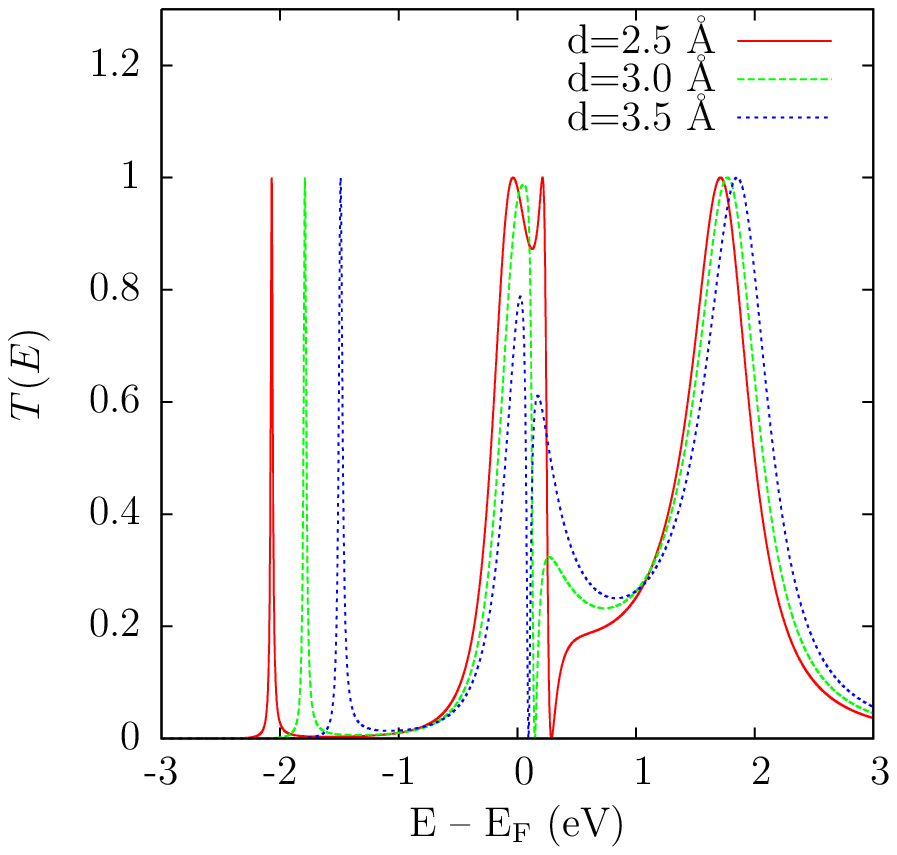}(b)\\
\includegraphics[width=0.45\textwidth,clip]{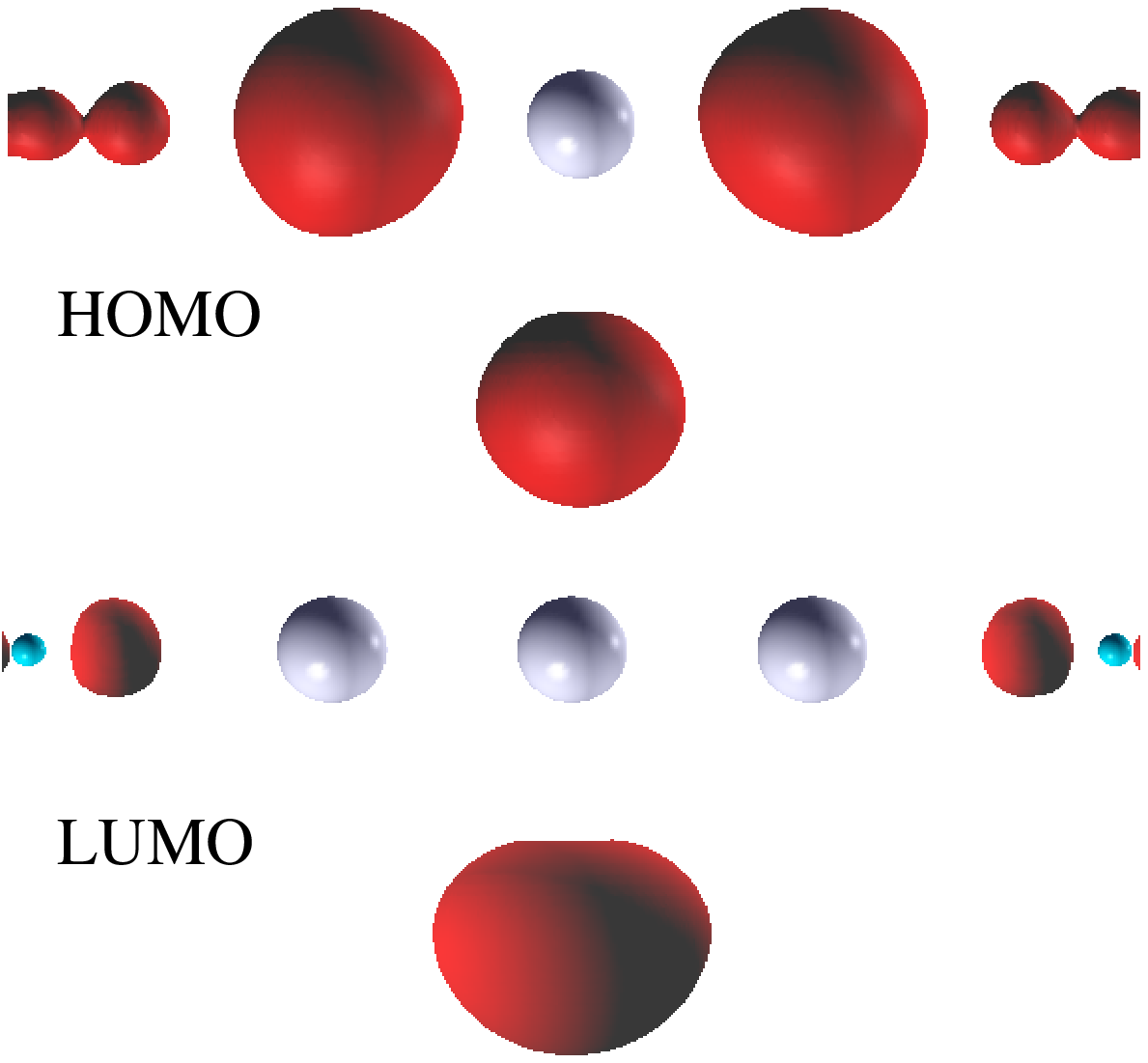}(c)
\includegraphics[width=0.45\textwidth,clip]{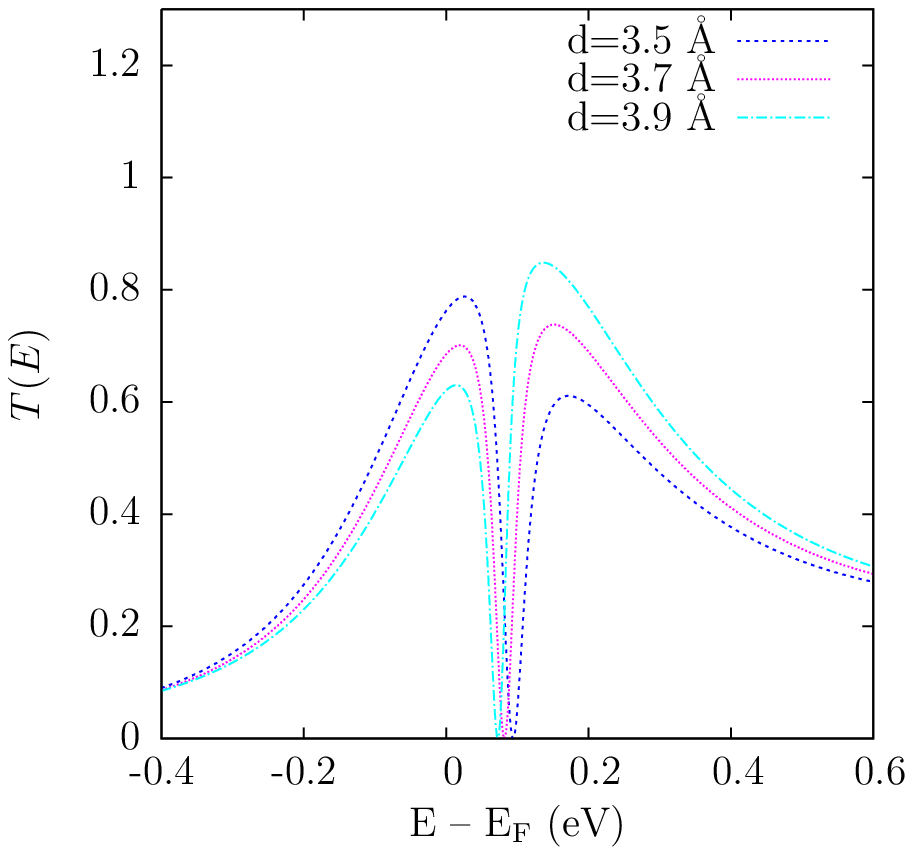}(d) 
\caption{(a) DOS (divided by 4, for easy comparison, in units of 1/eV)
and transmission coefficient of configuration (a) for $d=3.5$ \AA. (b) Transmission
coefficient for various distances between chain and impurity. (c) Charge density
iso-surfaces of the HOMO and the LUMO. (d) Zoom of (b) around $E_F$.}
\label{fig3}
\end{figure}

When a finite Li chain is coupled to H chain leads, charge is transfered between the
leads and the scattering region. For an even number of Li atoms this charge transfer results
in a level shift towards $E_F$ \cite{He07}. Extending the studies of \cite{He07}, 
we discuss in the following the influence of perturbations.
For configuration (a), with an additional Li atom adjacent to the Li chain, 
the tight-binding energy levels of the scatterer are given by 
$\varepsilon_{1,4}=\pm \sqrt{2t_{\rm par}^2+t^2_{\rm per}}$, and $\varepsilon_{2,3}=0$. 
The coupling to the leads lifts the degeneracy of the zero-energy level. 
A comparison of the density of states (DOS) obtained by SIESTA with $T(E)$ 
is given in Fig.\ \ref{fig3}(a) for a projection of the DOS onto the Li atoms. 
The distance of the ad-atom to the chain is $d=3.5$ \AA. The four energy levels of the scatterer
are located at $-1.5$ eV, 0 eV $= E_F$, 0.3 eV, and 1.7 eV.
The smaller peak at $-1.0$ eV as well as the shoulder at 1.6 eV arise from hybridization
with atoms in the leads. When the distance between the ad-atom and the Li chain is
increased, the splitting of the states near $E_F$ becomes more pronounced, see Fig.\
\ref{fig3}(b).

Analyzing the spatial distribution of the charge for each energy level yields
strong similarities between the tight-binding model and the DFT calculation.
The lowest level corresponds to a homogeneous charge distribution over the scatterer. 
The second level (highest occupied molecular orbital, HOMO) resides on Li atoms 1, 3, and 4,
whereas the third level (lowest unoccupied molecular orbital, LUMO) is dominated by
the ad-atom, see Fig.\ \ref{fig3}(c). In the range $d=3.5 \ldots 4.0$ \AA\ a
redistribution of weight from the HOMO to the LUMO is apparent.
For increasing distance $d$, the charge becomes more localized on the ad-atom,
resulting in a stronger transmission through the LUMO. The minimum of the total energy
appears for $d=2.75$ \AA; then the energy increases almost linearly when increasing $d$
(up to $d=4.0$ \AA).

\begin{figure}
\includegraphics[width=0.45\textwidth,clip]{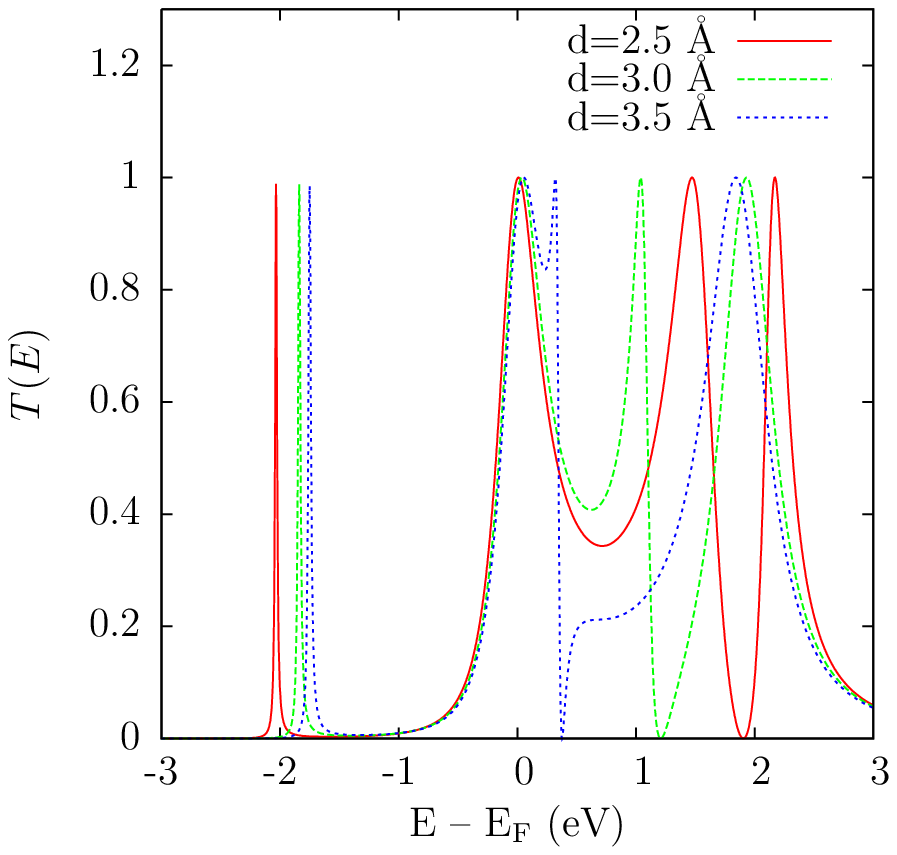}
\includegraphics[width=0.45\textwidth,clip]{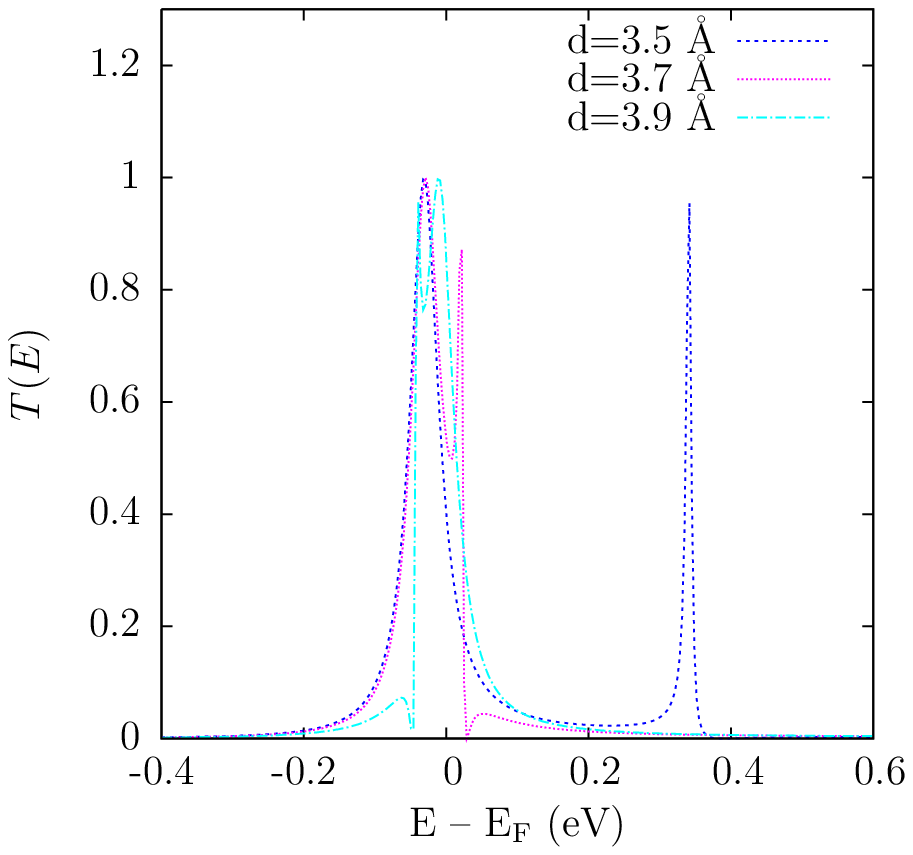}
\caption{Transmission coefficient for configuration (b) with strong ($a=2.8$ \AA, left) and
weak ($a=4.0$ \AA, right) coupling to the leads.}
\label{fig_confb}
\end{figure}

For the symmetrical configuration (b), see Fig.\ \ref{fig_confb}, we find five
energy levels in the tight-binding spectrum: bonding and antibonding levels 
$\varepsilon_{1,5}=\pm \sqrt{2t_{\rm par}^2+2t^2_{\rm per}}$ as well as $\varepsilon_{2,3,4}=0$.
A next-nearest-neighbor hopping $t_{\rm nn}$ shifts one of the zero-energy levels, say,
$\varepsilon_{4}$, to a finite value, given by
$\varepsilon_4\approx 2t_{\rm nn}t_{\rm per}/t_{\rm par}$
(which holds in the limit $t_{\rm nn}, t_{\rm per} \ll t_{\rm par}$). The coupling to the
leads also lifts the threefold degeneracy of $\varepsilon_{2,3,4}$. As a result, a spectrum of five 
levels is obtained, with two nearly degenerate levels near the Fermi energy,
see Fig.\ \ref{fig_confb}. Level 4, corresponding to $\varepsilon_4$, has a rather homogeneous charge
distribution over the scattering region for a small distance, $d=2.5$ \AA. With increasing distance,
the contribution of atom 2 to the charge density decreases, and hence the transmission through this
level.

The position of the fourth peak in $T(E)$
depends strongly on the distance between the ad-atoms and the chain. For $d=3.5$ \AA,
for example, it is located at about 0.5 eV, whereas it is found at 1.0 eV for $d=3.0$ \AA\
(equilibrium distance). The two peaks near $E_F$ cannot be resolved on the left hand side
of Fig.\ \ref{fig_confb} due to the coupling to the leads. For weak coupling a
three-peak structure around $E_F$ is resolved, again showing the redistribution of weight
below and above $E_F$, compare Fig.\ \ref{fig3}(d) with the right hand side of Fig.\ \ref{fig_confb}.

\begin{figure}
\includegraphics[width=0.45\textwidth,clip]{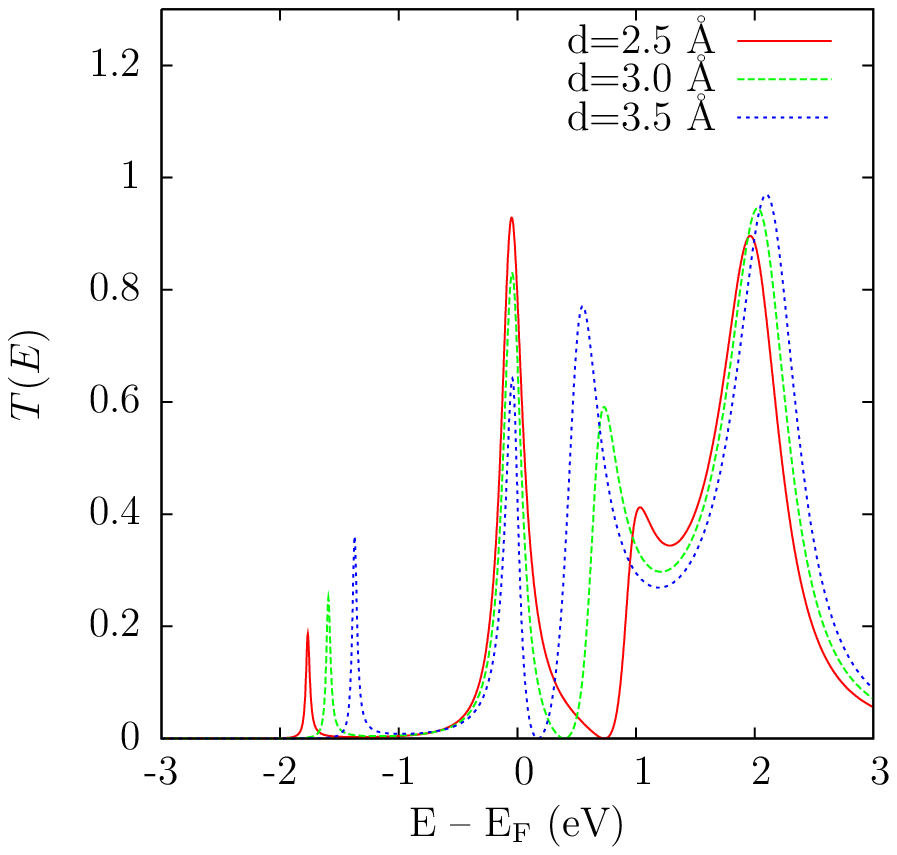}
\includegraphics[width=0.45\textwidth,clip]{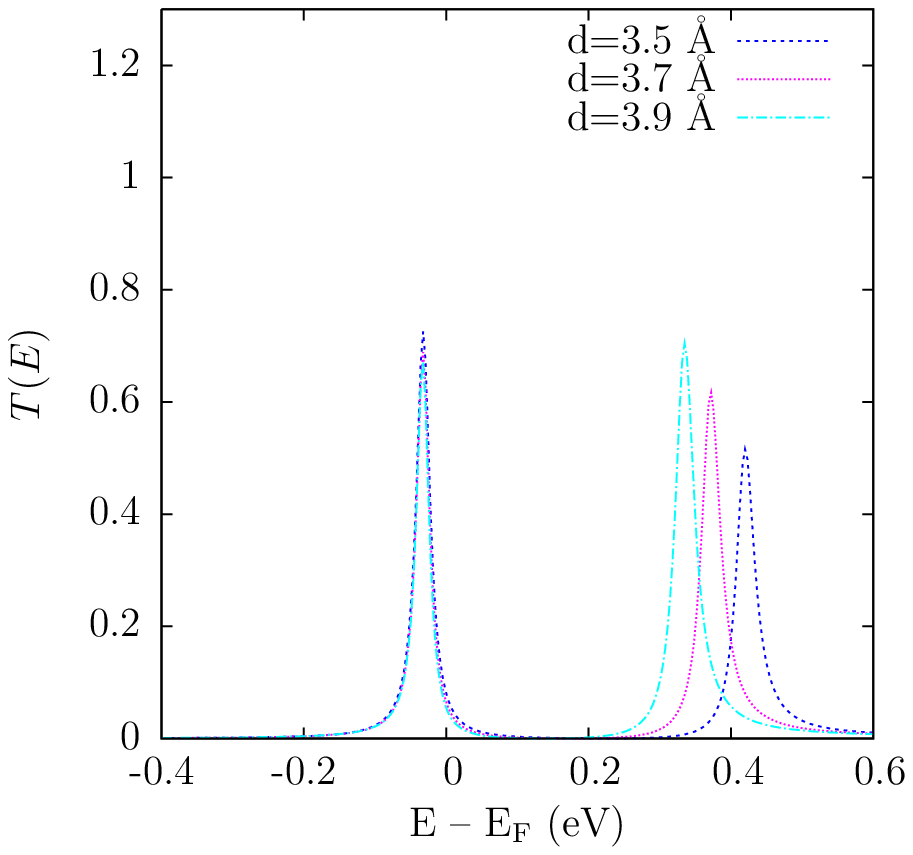}
\caption{Transmission coefficient for configuration (c) with strong ($a=2.8$ \AA, left) and
weak ($a=4.0$ \AA, right) coupling to the leads.}
\label{fig5}
\end{figure}

Configuration (c) corresponds to a scatterer which is asymmetric in both in-plane
directions. The tight-binding model yields the spectrum
$\varepsilon_{1,4}\approx \pm \sqrt{2}t_{\rm par}$ and
$\varepsilon_{2,3}\approx \pm t_{\rm s}/\sqrt{2}$ provided $t_{\rm s}$ is small
$t_{\rm s} \ll t_{\rm par}$, i.e., when the distance between ad-atom and chain is large.
The first guess thus is that the system should be insulating,
in contrast to the parent chain and to the
configuration where the ad-atom is attached to a single chain atom.
According to Fig.\ \ref{fig5}, however, $T(E)$ does not show a minimum at $E_F$: in fact,
due to the charge transfer with the leads the HOMO is fixed near $E_F$, as demonstrated
in \cite{He07} (also within DFT). In addition, the lowest level shifts slightly
in energy when varying $d$.

Note that for $d=2.5$ \AA\ the distance between ad-atom and chain atoms is only about
$d=2.9$ \AA\, hence smaller than the intra-chain nearest-neighbor distance of $d=3.0$ {\AA}.
Nevertheless, $T(E)$ for $d=2.5$ \AA\ essentially is the same as the transmission for larger
distances. For $d$ above 3.0 \AA\, the HOMO and its transmission peak remain unchanged;
however, the LUMO shifts to lower energy with increasing $d$ (i.e., for decreasing $t_{\rm s}$).

\begin{figure}
\includegraphics[width=0.45\textwidth,clip]{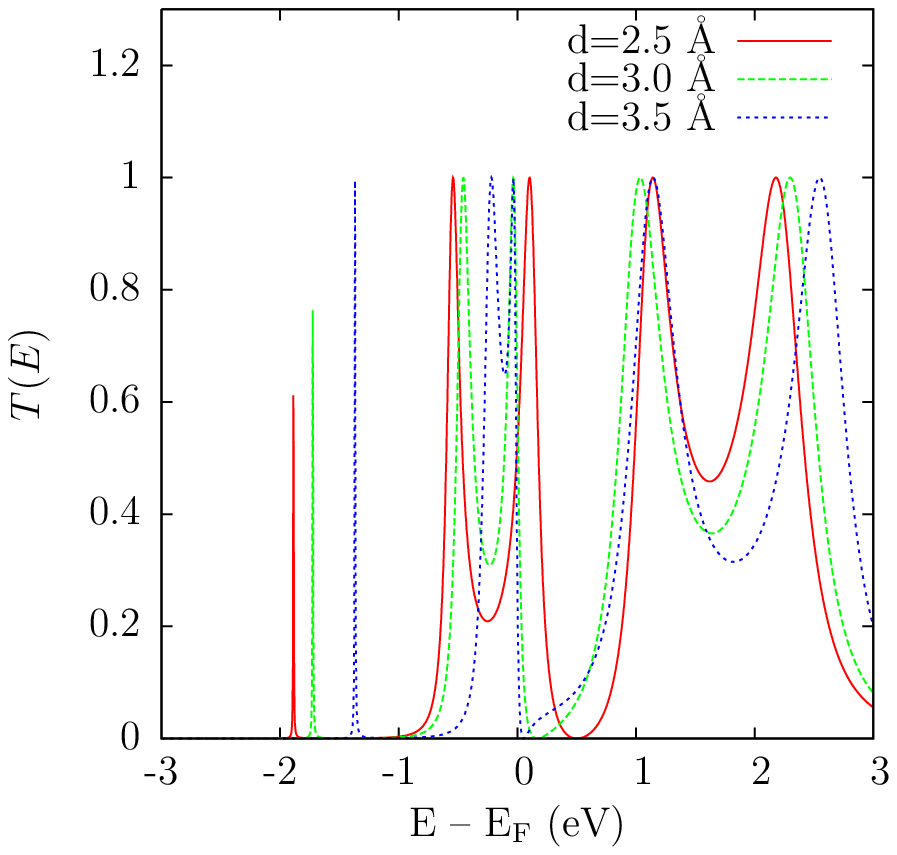}
\includegraphics[width=0.45\textwidth,clip]{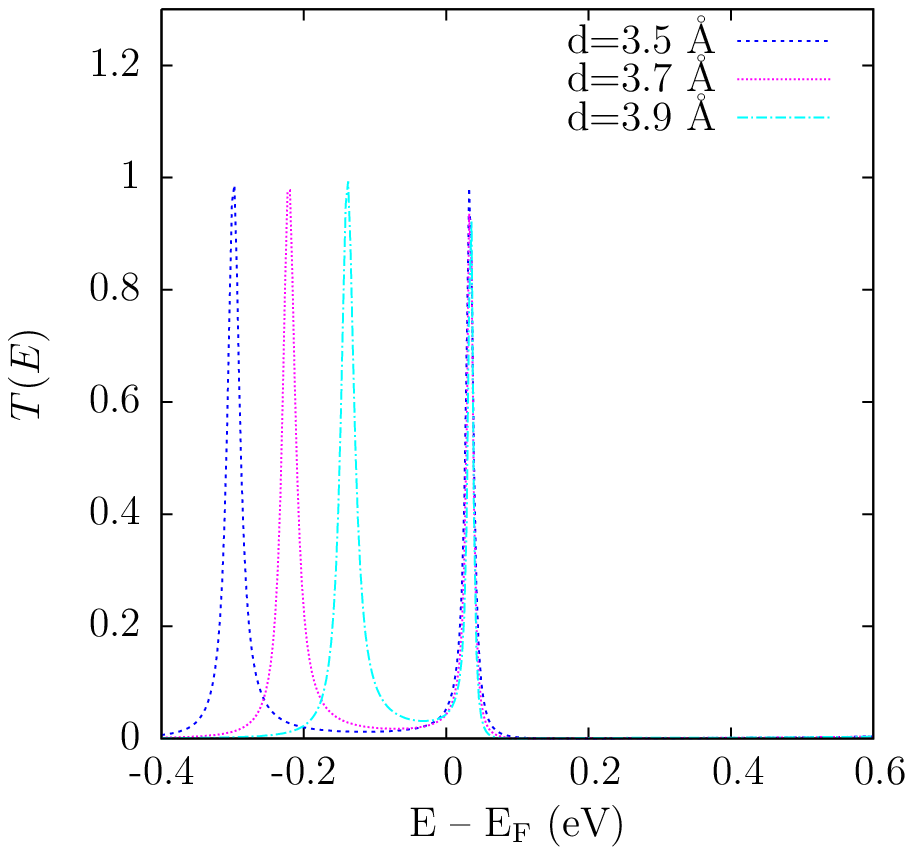}
\caption{Transmission coefficient for configuration (d) with strong ($a=2.8$ \AA, left) and
weak ($a=4.0$ \AA, right) coupling to the leads.}
\label{fig6}
\end{figure}

Symmetry in the transport direction can be obtained by placing the ad-atom below the
central bond of a four-atom Li chain, see configuration (d). In this case the parent chain
is insulating as $\varepsilon_{1,2,3,4}=(\pm 1/2\pm\sqrt{5}/2) t_{\rm par}$. Including the
impurity yields $\varepsilon_{1,4}=-t_{\rm par}/2\pm\sqrt{5t_{\rm par}^2+8t_{\rm s}^2}/2$, 
$\varepsilon_{2,5}=t_{\rm par}/2\pm\sqrt{5}t_{\rm par}/2$, and $\varepsilon_{3}=0$. 
Thus, the ad-atom again switches the electronic state of the chain, but now from insulating
to metallic. If we couple the central chain to the leads, $\varepsilon_{3}$ is shifted to
higher energy, due to charge transfer, see Fig.\ \ref{fig6}. For $d>3.0$ \AA\ the
energy of the LUMO and its transmission peak remain essentially unaltered, but the distance
to the HOMO decreases as $t_{\rm s}$ decreases, similar to the behavior of configuration (c).

\begin{figure}
\includegraphics[width=0.45\textwidth,clip]{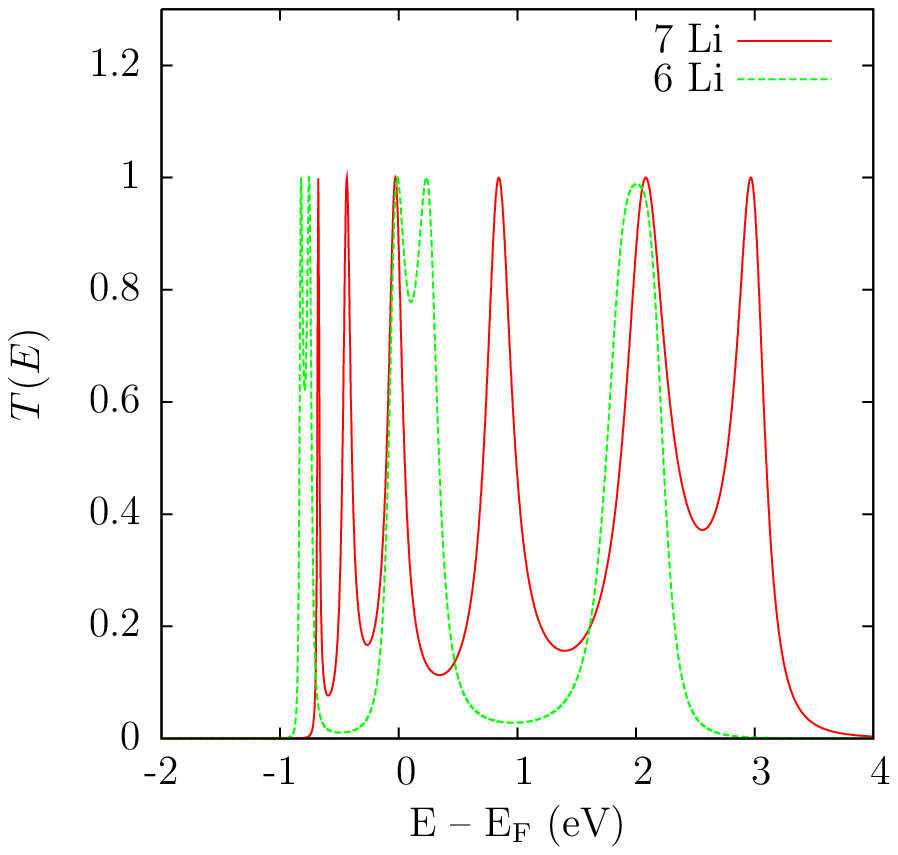}
\includegraphics[width=0.45\textwidth,clip]{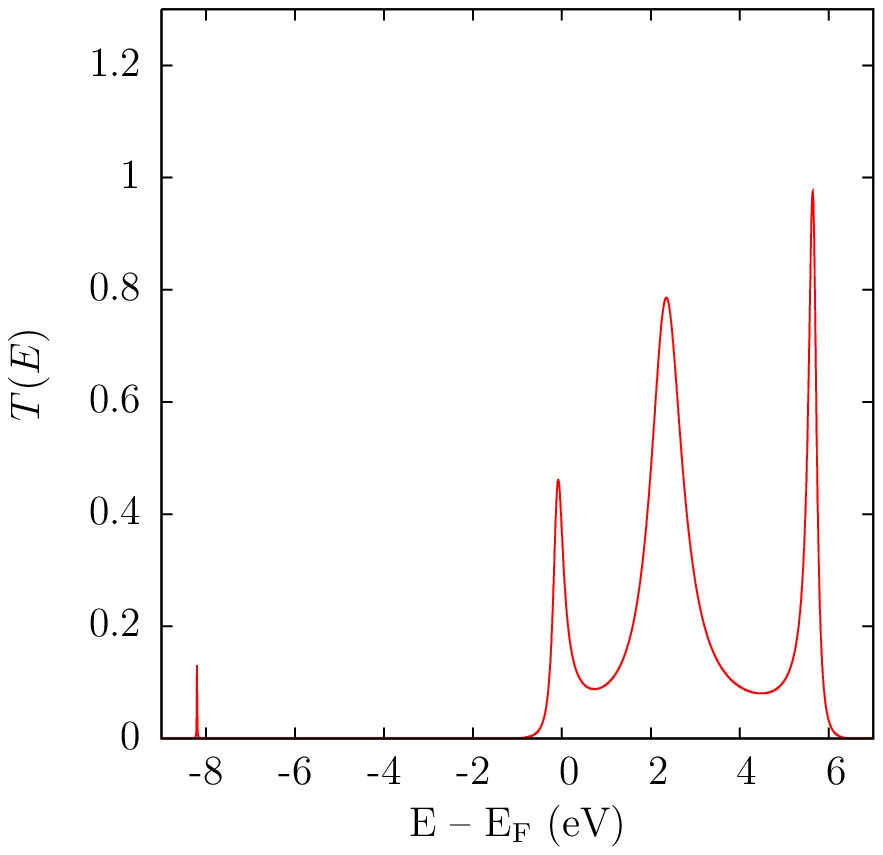}
\caption{Left: Transmission coefficient for a seven-atom Li chain with an F ad-atom attached
directly to the central Li atom, and for a six-atom Li chain with weak link at the center.
Right: Transmission coefficient for a Li chain with intra-chain F impurity.}
\label{fig7}
\end{figure}

We next study the influence of an impurity different from the atoms in the central region,
comparing ad-atom and intra-chain configurations.
As a prototypical example with high electronegativity we choose a fluor atom, which we place
either adjacent to or incorporated into the Li chain. In the former case we
expect, for an F atom attached to a single chain atom, that one Li electron will be trapped in
the Li-F cluster, no longer contributing to transport. According to the left
hand side of Fig.\ \ref{fig7}, the transmission spectrum of a seven-atom
Li chain with F ad-atom in the middle ($d=2$ \AA\ away from the chain) consists of six peaks,
confirming the above picture of one trapped electron.
We note that the chain remains metallic despite the impurity. This can be understood
by studying an equivalent system with a six-atom Li chain and a weak link at the center.
To model the weak link, we increase the distance between the two central Li
atoms to 5 \AA. Effectively, we obtain the transmission spectrum of two three-atom Li chains,
see the dashed curve on the left hand side of Fig.\ \ref{fig7}, i.e., metallic behavior and a
high transmission at $E_F$. 

The situation is different when the electronegative atom is placed within the chain, 
as shown on the right hand side of Fig.\ \ref{fig7} for a Li-Li-F-Li configuration with
a Li-F distance of 1.5 \AA. We find that the F atom forms a site with large
attractive potential, since one transmission peak appears at very low energy (at about $-8.2$ eV).
We note that the resonance level at $E_F$ does not attain a transmission of 1.
While an F ad-atom forms a neutral cluster with the adjacent Li atom, an intra-chain F atom
couples to both Li neighbors and acts as strong attractive potential.

\section{Conclusion}

We have investigated the influence of impurities on the transmission of metallic
and insulating monovalent Li chains. We have demonstrated that
an impurity connected to a single atom of the chain leads to a splitting of the zero-energy
peak, causing a strong dependence of the conductance on structural details of the scatterer. 
An impurity connected to two atoms of the chain with equal distance, on the other hand,
switches the system from insulating to metallic behavior, or vice versa.
The transmission coefficient, however, in both cases shows a minimum at $E_F$ as a consequence
of the charge transfer with the leads. Charge transfer with the leads prohibits a
prediction of details of the transmission spectrum by tight-binding models of the scatterer. 
Electronegative ad-atoms localize electrons in the chain but do not decouple it, whereas
electronegative impurities within the chain result in low-lying transmitting levels and,
thus, shift the spectrum to higher energy. Our results demonstrate that
switches based on mono-atomic chains show a high variability of possible modifications
that can be used to tailor the switching behavior. 

\begin{acknowledgments}
We gratefully acknowledge discussions with  S.\ Schenk and P.\ Schwab, and
thank the Deutsche Forschungsgemeinschaft (TRR 80) for financial support.
\end{acknowledgments}


\begin{thebibliography}{10}
\bibitem{note} Note that we use ``central region'', ``scattering region'', and ``scatterer'' synomymously.
\end{thebibliography}
\end{document}